\documentclass[12pt]{article}
\usepackage{graphicx}
\usepackage{subfigure}
\usepackage{amsmath,amsthm}
\usepackage{fullpage}
\begin{document}
\begin{center}
{\textbf{ \large Effect of dust-ion collision on superthermal plasmas in cylindrical and spherical geometry }}
\end{center}{\large\par}
\vspace*{.0in}
	\begin{center}
		{\textbf{Uday Narayan Ghosh$^{1}$ Laxmikanta Mandi$^{2,3}$ and Prasanta Chatterjee$^{3}$ } \vspace*{.25in}\\
			\textbf{$^{1}$Department of Mathematics,KKM College,\\Jamui, Bihar 811307, India }\\
		\textbf{$^{2}$Department of Mathematics,University of Gour Banga,\\Malda, West Bengal 732101, India }\\
		\textbf{$^{3}$Department of Mathematics, Siksha Bhavana,}\\ \textbf{Visva Bharati University, Santiniketan-731235, India}}\\
		
		\vspace*{.1in} Contributing Author $^{1}$ E-mail:unghosh@gmail.com \\
		\vspace*{.1in} Corresponding Author $^{2,3}$ E-mail: laxmikanta11@gmail.com\\
		\vspace*{.1in} Contributing Author $^{3}$ E-mail: prasantachatterjee1@rediffmail.com
	\end{center}
\begin{abstract}
Dust-ion collisional effect in dust ion acoustic waves is investigated in the framework of spherical and cylindrical geometry. For such study the fluid equation are takes and collisional effect is consider in the equation of momentum balance. Using reductive perturbation technique the evolution equation is derived with additional damping term for non planner geometry. Using the conservation principle an analytical time dependent solitary wave solution is obtained for the evolution equation. It is seen that dust ion collisional frequency has an significant effect on the width and amplitude of the solitary wave solution.    
\end{abstract}
\section{Introduction}
Presence of macroscopic dimensional dust particles, compared to atoms and ionized nuclei are naturally of the order of a micron. They are mainly carbonaceous, silicate, or ferrite in nature, or an alloy of these. In the astrophysical environment, it can be greatly varied with their compositions. Dust grains are also coming from pie of macromolecules and frozen ices. The fascinating research area has been opened up due the interplay between plasmas and charged dust grains. The mixtures of ordinary plasma particles and charged dust grains, has given birth the new subject named dusty plasmas. In many solar systems and interstellar environments\cite{d1}-\cite{d2}, dusty plasma is too much nearer to the reality. Due to the interactions of dust grains with the ordinary plasmas, considerable differences are been encountered. Dusty plasma generates various types of new collective modes within their ultra-low-frequency regime, that do not exist in the usual electron-ion plasmas.
As the dust grains are billion times heavier than usual charge particles, they are surrounded by ionized particle in plasma environments and becomes highly charged. Thus they bring  important contributions to plasmas arising from electromagnetic interactions .So, dusty plasma consequently has received wide attention in recent years\cite{d4}-\cite{di20}.  
The mostly investigated modes are dust acoustic mode\cite {d10}, dust drift mode\cite {t9}, Shukla-Varma mode\cite {t10}, dust lattice mode\cite {di15}, dust cyclotron mode\cite {t12}, dust ion acoustic (DIA) mode\cite {t13}, and dust Berstain-Green-Kruskal mode\cite {t14}.
Theoretical investigations and experimental observations recognise that the dust charge dynamics introduces new episode \cite{d10}-\cite{d14} where dust particle mass provides the inertia and the pressures of electrons and ions give rise to the restoring force. Rao et al. \cite{d10} first predicted theoretically the existence of extremely low-phase-velocity (in comparison with the electron and ion thermal velocities) dust acoustic waves(DIAWs) in an unmagnetized dusty plasma. The laboratory experiments \cite{d13}-\cite{d14} have conclusively verified this theoretical prediction and have shown some nonlinear features of DIAWs.\\
The charged dust grains can modify the propagation of IAWs, even through the dust grains do not participate in the wave dynamics. Accordingly a new mode DIAWs were predicted by Shukla and Silin \cite{di13}. For DIAWs the phase velocity is much smaller (larger) than the electron thermal speed (ion and dust thermal speeds). As the dust grains has large mass, the ion plasma frequency $\omega _{pi}$ is much larger than the dust plasma frequency $\omega _{pd}$. This shows that the phase velocity of the DIAWs in a dusty plasma is larger than $c_s=\sqrt{k_BT_e/m_i}$. The DIAWs have been observed in laboratory experiments by Barkan et al \cite{ka14}, Nakamura et al \cite{ka15} and Melrino et al \cite{ka16} by means of a grid inserted into the plasma column produced in a Q-machine. The parameters are (by Barken et al \cite{ka14}) is $n_i = n_e= 10^^6 cm^{-3}$, $T_e\cong T_i\cong 2320$K, $r_d\cong 5\mu m$ and $m_d=10^{-9}$g. It is shown that the frequencies of the DIA waves for laboratory plasma parameters are order of kilo Hz. Which are consistent with the theoretical prediction of Shukla and Silin \cite{di13}. \\
The electromagnetic interaction can significantly influence the collective dynamics of the plasma specially in the presence of dust-ion collision\cite{dm1}-\cite{dm5}. 
Choudhury et al \cite{dm1} recently studied the nature of solitary waves structures in a quantum semiconductor plasma in presence of electron-phonon collision frequency. They have derived damped KdV equations from fluid model and investigated the effects of parameters in different semiconductors. Mainly they have shown the damping effect due to collision plays a very important role upon the structures of solitary waves. 
Most recently, Das et al.\cite{dm5} have investigated the dynamics of dust ion acoustic waves (DIAWs)in a magnetized dusty plasma with superthermal electrons in the framework of a damped Zakharov-Kuznetsov (dZK) equation. They have investigated chaotic motion with the help of the quasiperiodic route to chaos, and shown that the parameter collision frequency $v_{id0}$ plays the role as a switching parameter from the quasiperiodic route to chaos for the DIAWs.\\
Most of these investigations \cite{d4}-\cite{di20} are confined to unbounded planar geometry, which may not be a realistic situation in laboratory devices and space. Recently, theoretical studies \cite{n1}-\cite{n4} show that the properties of solitary waves in bounded nonplanar cylindrical/spherical geometry are very differ from those in unbounded planar geometry and therefore a great deal of attention is paid to understand the nonlinear phenomena like solitons, shocks, interactions of waves etc. \cite{n5}-\cite{n8} in the nonplanar geometry.\\
Ghosh et al.\cite{dn1} have investigated head-on collisions between dust-ion acoustic solitary waves by employing extended version of Poincar'e-Lighthill-Kuo perturbation method, for a plasma having stationary dust grains, inertial ions, and nonextensive electrons in the nonplanar geometery. They have shown that the nonplanar geometry modified analytical phase-shift after a head-on collision is derived.\\
The paper is composed in the following manner: In Sec.$2$ we have discussed the model equation and the superthermal electron distribution. The derivation of nonplanner damped KdV equation and determination of its approximate analytical solitary wave solution has been shown in the Sec.$3$. Numerical analysis and discussion have been illuminated in the Sec.$4$ and Sec.$5$ is kept for conclusion. 

\section{Basic Equations}
In this work, we consider an unmagnetized collisional dusty plasma that contains cold inertial ions, stationary dusts with negative charge and non-inertial $\kappa$-distributed electrons. The normalized ion fluid equations which include the equation of continuity, equation of momentum balance and Poisson equation, governing the DIA waves, are given by
\begin{eqnarray}
\frac{\partial n}{\partial t}+\frac{1}{r^m}\frac{\partial (r^mnu)}{\partial r}&=&0,\label{p4eq2}\\
\frac{\partial u}{\partial t}+u\frac{\partial u}{\partial r}&=&-\frac{\partial \phi}{\partial r}-\nu_{id} u,\label{p4eq3}\\
\frac{1}{r^m}\frac{\partial}{\partial r}(r^m\frac{\partial\phi}{\partial r})&=&(1-\mu)n_e-n+\mu.\label{p4eq4}
 \end{eqnarray}

where $n$ is the number density of ions normalized to its equilibrium value $n_{0}$, $u$ is the ion fluid velocity normalized to
ion acoustic speed $C_{s}=\sqrt(\frac{k_BT_e}{m_i})$, with $T_e$ as electron temperature, $k_B$ as Boltzmann constant and
$m_i$ as mass of ions. The electrostatic wave potential $\phi$ is normalized to $\frac{k_B T_e}{e}$, with $e$ as magnitude of electron charge.
The space variable $x$ is normalized to the Debye length $\lambda_D=(\frac{T_e}{4{\pi}n_{e0} e^2})^\frac{1}{2}.$ and the time $t$
is normalized to $\omega_{pi}^{-1}=(\frac{m_i}{4{\pi}n_{e0} e^2})^\frac{1}{2}$, with $\omega_{pi}$ as ion-plasma frequency.
Here $\nu_{id}$ is the dust-ion collisional frequency and $\mu=\frac{Z_d n_{d0}}{n_0}$.
Most of space/astrophysical and laboratory environments confirm the presence of superthermal electrons and ions\cite{k7}-\cite{k9}. Superthermal electrons and ions are modeled by a kappa type distribution rather than Maxwellian.
In order to describe superthermal electron, we assumed that they followed the isotropic $\kappa$ velocity distribution given as\cite{Vasyliunas1968} 
\begin{eqnarray}
f_{\kappa}(v)=\frac{n_{e0}}{\pi^{3/2}\theta^3}\frac{\Gamma(\kappa+1)}{\sqrt{\kappa}\Gamma(\kappa-1/2)}\Big(1+\frac{v^2}{\kappa\theta^2}\Big),\label{kappa}
\end{eqnarray}
where $n_{e0}$ is the equilibrium electron density,$\kappa$ is a parameter representing the spectral 
index of the distribution, $\Gamma(\kappa)$  is the gamma function and $\theta$ is a modified thermal 
speed given by $\theta^2=(\frac{\kappa-3/2}{\kappa})(\frac{2k_BT}{m_e})$ with the spectral index $\kappa>3/2$. Here $T$ and $m_e$ represent the electron temperature and electron mass, 
respectively. When $\kappa\rightarrow\infty$, the kappa-distribution function given by (\ref{kappa}) 
reduces to a Maxwellian distribution.
The normalized superthermal electron number density  is given by
\begin{eqnarray*}
n_e=\bigg(1-\frac{\phi}{\kappa-3/2}\bigg)^{-\kappa+1/2}.
\end{eqnarray*}
\section{Nonlinear Analysis}
 To study the nonlinear propagation of DIA waves in a collisional dusty plasma, the reductive perturbation technique (RPT) is applied  to derive the non-planner damped KdV equation. According to RPT the independent variables are stretched as
 \begin{eqnarray}
 \xi&=&\epsilon^{1/2}(r-vt),\label{p4eq5}\\
 \tau&=&\epsilon^{3/2}t \label{p4eq6}
 \end{eqnarray}
 where $\epsilon$ is the strength of nonlinearity and $v$ be the phase velocity of the DIA waves. The expansions of the  dependent variables
 are as follows:
 \begin{eqnarray}
 n=1+\epsilon n_1+\epsilon^2n_2+\cdot \cdot\cdot,\label{p4eq7}\\
 u=0+\epsilon u_1+\epsilon^2u_2+\cdot \cdot\cdot,\label{p4eq8}\\
 \phi=0+\epsilon \phi_1+\epsilon^2\phi_2+\cdot \cdot\cdot,\label{p4eq9}\\
 \nu_{id}\sim\epsilon^{3/2}\nu_{id0}.\label{p4eq10}
 \end{eqnarray}
Substituting the above expansions along with stretching coordinates into equations (\ref{p4eq2})-(\ref{p4eq4}) and equating the coefficients of lowest order
of $\epsilon$, the dispersion relation is obtained as
\begin{eqnarray}
v=\frac{1}{\sqrt{a(1-\mu)}},\label{p4eq11}
\end{eqnarray}
with $a=\frac{\kappa-1/2}{\kappa-3/2}$.\\
Taking the coefficients of next higher order of $\epsilon$, we obtain the non-planner damped KdV equation
\begin{equation}
 \frac{\partial \phi_1}{\partial \tau}+\frac{\gamma}{2\tau}\phi_1+A\phi_1\frac{\partial \phi_1}{\partial \xi}+B\frac{\partial^3 \phi_1}{\partial \xi^3}+C\phi_1=0, \label{p4eq12}
\end{equation}
where $A=\bigg(\frac{3-2b(1-\mu)v^4}{2v}\bigg)$, $B=\frac{v^3}{2}$
and $C=\frac{\nu_{id0}}{2}$, with $b=\frac{\kappa^2-1/4}{2(\kappa-3/2)^2}.$\\

In absence of damping and non-planner effect \textit{i.e.,} for $C+\frac{\gamma}{2\tau}=0$ the Eq.(\ref{p4eq12}) takes the form of well known KdV equation with the solitary wave solution
\begin{equation}\label{p4eq14}
 \phi_1=\phi_m sech^2\bigg(\frac{\xi-M\tau}{W}\bigg),
\end{equation}
where $\phi_m=\frac{3M}{A}$ and $W=2\sqrt{\frac{B}{M}}$, with $M$ as the Mach number.\\
In this case, it is well established that
\begin{equation}
I=\int_{-\infty}^{\infty} \phi_1^2~d\xi,\label{p4eq15}
\end{equation}
is a conserved quantity.
For small values of $(C+\frac{\gamma}{2\tau})$, let us assume that the solution of equation (\ref{p4eq12}) is of the form
\begin{equation}\label{p4eq16}
\phi_1=\phi_m(\tau)sech^2 \bigg(\frac{x-M(\tau)\tau}{W(\tau)}\bigg),
\end{equation}
where $M(\tau)$ is an unknown function of $\tau$ and $\phi_m(\tau)=\frac{3M(\tau)}{A}$, $W(\tau)=2\sqrt{B/M(\tau)}$.\\
Differentiating equation (\ref{p4eq15}) with respect to $\tau$ and using equation (\ref{p4eq12}), one can obtain
\begin{eqnarray}
\frac{dI}{d\tau}+2(C+\frac{\gamma}{2\tau})I=0.\label{p4eq17}
\end{eqnarray}
Again,
\begin{eqnarray}
I&=&\int_{-\infty}^{\infty} \phi_1^2~d\xi,\label{p4eq19}\\
I&=&\int_{-\infty}^{\infty} {\phi_m}^2(\tau)sech^4\bigg(\frac{\xi-M(\tau)\tau}{W(\tau)}\bigg)~d\xi,\label{p4eq20}\\
I&=&\frac{24\sqrt{B}}{A^2}M^{3/2}(\tau).\label{p4eq21}
\end{eqnarray}
From equations (\ref{p4eq17}) and (\ref{p4eq21})we get \\
\begin{eqnarray}
\frac{dM(\tau)}{d\tau}=-\frac{4}{3}\big(\frac{\gamma}{2\tau}+C\big)M(\tau)\\
\Rightarrow M(\tau)=M_0\big(\frac{\tau_0}{\tau}\big)^{-\frac{2\gamma}{3}}e^{\frac{4C}{3}(\tau_0-\tau)},
\end{eqnarray}
where $M(\tau)=M_0$ at $\tau=\tau_0$.\\
Therefore, the analytical solitary wave solution of the dust ion acoustic waves for the non-planner damped nonplanar KdV equation (\ref{p4eq12}) is
\begin{equation}\label{p4eq22}
 \phi_1=\phi_m(\tau)sech^2\bigg(\frac{\xi-M(\tau)\tau}{W(\tau)}\bigg),
\end{equation}
where $\phi_m(\tau)=\frac{3M(\tau)}{A}$ and $W(\tau)=2\sqrt{\frac{B}{M(\tau)}}$.
\section{Effects of parameters and discussions}
Though out the study of plasma, it has been seen that the study of plasma waves are done in the framework of nonlinear evolution equation in planner or non-planner geometry. Most of these are regarding the derivation of cylindrical and spherical KdV equation and interactions of cylindrical and spherical solitary waves or damped KdV equations in planar geometry. Most of these investigations are carried out from the normalised fluid model. Where they have not incorporated the collisional effect of plasma species into the fluid model. It is quite natural to take account the collision effects of plasma species into the fundamental plasma fluid model and can not be ignored. The collision effect $\nu _{dn0} u$ of dust and ions are incorporated ignoring the collisional effect due to lighter mass of electrons compared to dust grains and ions.
In the planner geometry the effect of dust ion collision is incorporated as an extra term with the KdV type equation and named as damped KdV equation or so. In case of nonplanner geometry the collisional effect is not consider earlier. But in this article, we have derived this nonplanar damped KdV equation and  show various parameter impacts on the solitary waves structure and categorically signify our objectives. \\

In this section, we present the effects of different physical parameters $\kappa$ and $\nu_{id0}$ on the DIA solitary wave solution of the non-planner damped KdV equation (\ref{p4eq12}) through numerical computations. \\
\begin{figure}
\centering
\includegraphics[scale=0.5]{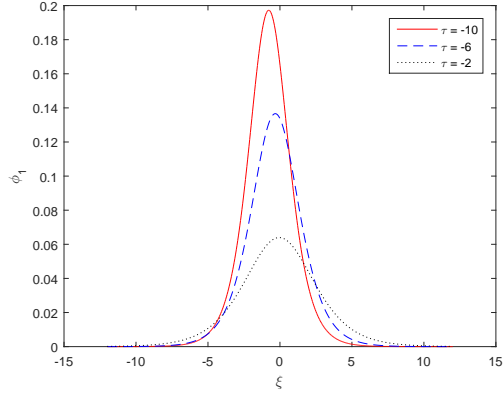}
\caption{Variation of the solitary wave of the non-planner damped KdV equation (\ref{p4eq12}) for different values of $\tau$ with $M=0.1,\kappa=1.8,\mu=0.1,\nu_{id0}=0.01,\tau_0=-14$ and $\gamma=1$.}\label{f1}
\end{figure}
\begin{figure}
\centering
\includegraphics[scale=0.5]{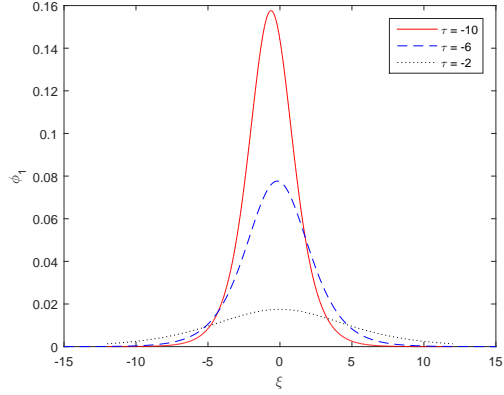}
\caption{Variation of the solitary wave of the non-planner damped KdV equation (\ref{p4eq12}) for different values of $\tau$ with $M=0.1,\kappa=1.8,\mu=0.1,\nu_{id0}=0.01,\tau_0=-14$ and $\gamma=2$.}\label{f2}
\end{figure}
In figure 1 and figure 2, variations of the solitary wave of the non-planner damped KdV equation (\ref{p4eq12}) with different values of $\tau$ are plotted for cylindrical and spherical geometry respectively. Other parameters are $M=0.1,\kappa=1.8,\mu=0.1,\nu_{id0}=0.01,\tau_0=-14$. These figures show the variation of electrostatics potential features are changing with the variation of the time evolved in that solitary wave profile. It is seen that amplitude of solitary is reduced in both cylindrical and spherical cases when $\tau $ is taking its higher magnitude. Therefore, the compressive solitary wave becomes flatter as the time increases. Thus, the compressive solitary wave solution of the nonplanar DKdV (13) becomes flattened enough with the time. It is well known that the prominent balance between nonlinearity and dispersion generates the particular shape of solitary waves while studying KdV equation in planar geometry without damping. Here damping is been incorporated in nonplanar geometry, simultaneous effect of damping and nonplanar. This is the main cause to modify the shape of the solitary waves profile of nonplanar damped KdV equation. Gradually, we will give indications for fulfilment of our findings. \\
\begin{figure}
\centering
\includegraphics[scale=0.5]{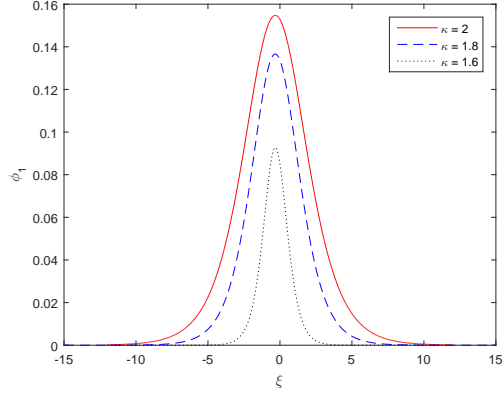}
\caption{Variation of the solitary wave of the non-planner damped KdV equation (\ref{p4eq12}) for different values of $\kappa$ with $M=0.1,\tau=-7,\mu=0.1,\nu_{id0}=0.01,\tau_0=-14$ and $\gamma=1$.}\label{f3}
\end{figure}

\begin{figure}
\centering
\includegraphics[scale=0.5]{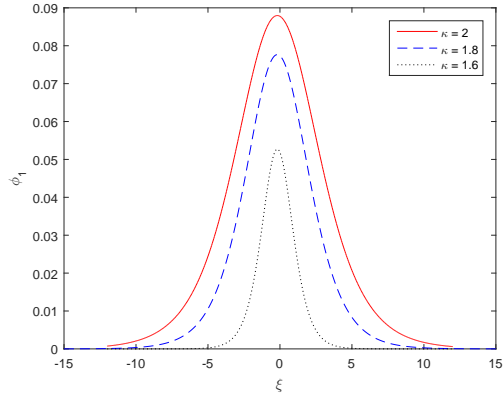}
\caption{Variation of the solitary wave of the non-planner damped KdV equation (\ref{p4eq12}) for different values of $\kappa$ with $M=0.1,\tau=-7,\mu=0.1,\nu_{id0}=0.01,\tau_0=-14$ and $\gamma=2$.}\label{f4}
\end{figure}
Variation of the solitary wave of the non-planner damped KdV equation (\ref{p4eq12}) for different values of $\kappa$ are plotted in figure 3 and figure 4 for  cylindrical and spherical geometry. It is depicted that the solitary wave profiles are changed significantly with $\kappa$. So superthermal parameter has keen effect upon the solitary  wave solutions. It is also seen that the increase of superthermal index diminishes the amplitude as well as the width of the solitary wave structures. The superthermal index $\kappa$ plays important role to change the solitary wave structure. \\
\begin{figure}
\centering
\includegraphics[scale=0.5]{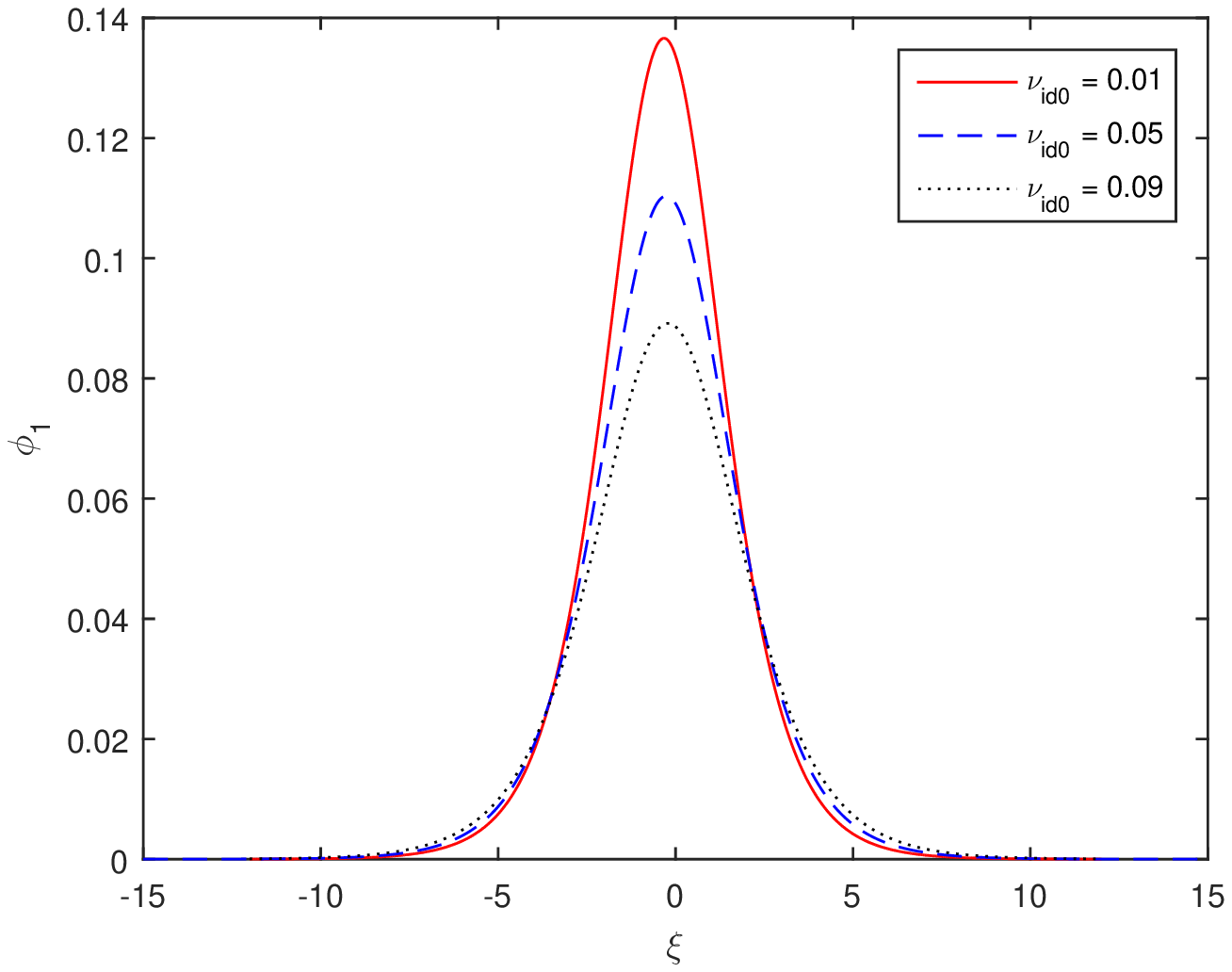}
\caption{Variation of the solitary wave of the non-planner damped KdV equation (\ref{p4eq12}) for different values of $\nu_{id0}$ with $M=0.1,\tau=-7,\mu=0.1,\kappa=1.8,\tau_0=-14$ and $\gamma=1$.}\label{f5}
\end{figure}

\begin{figure}
\centering
\includegraphics[scale=0.4]{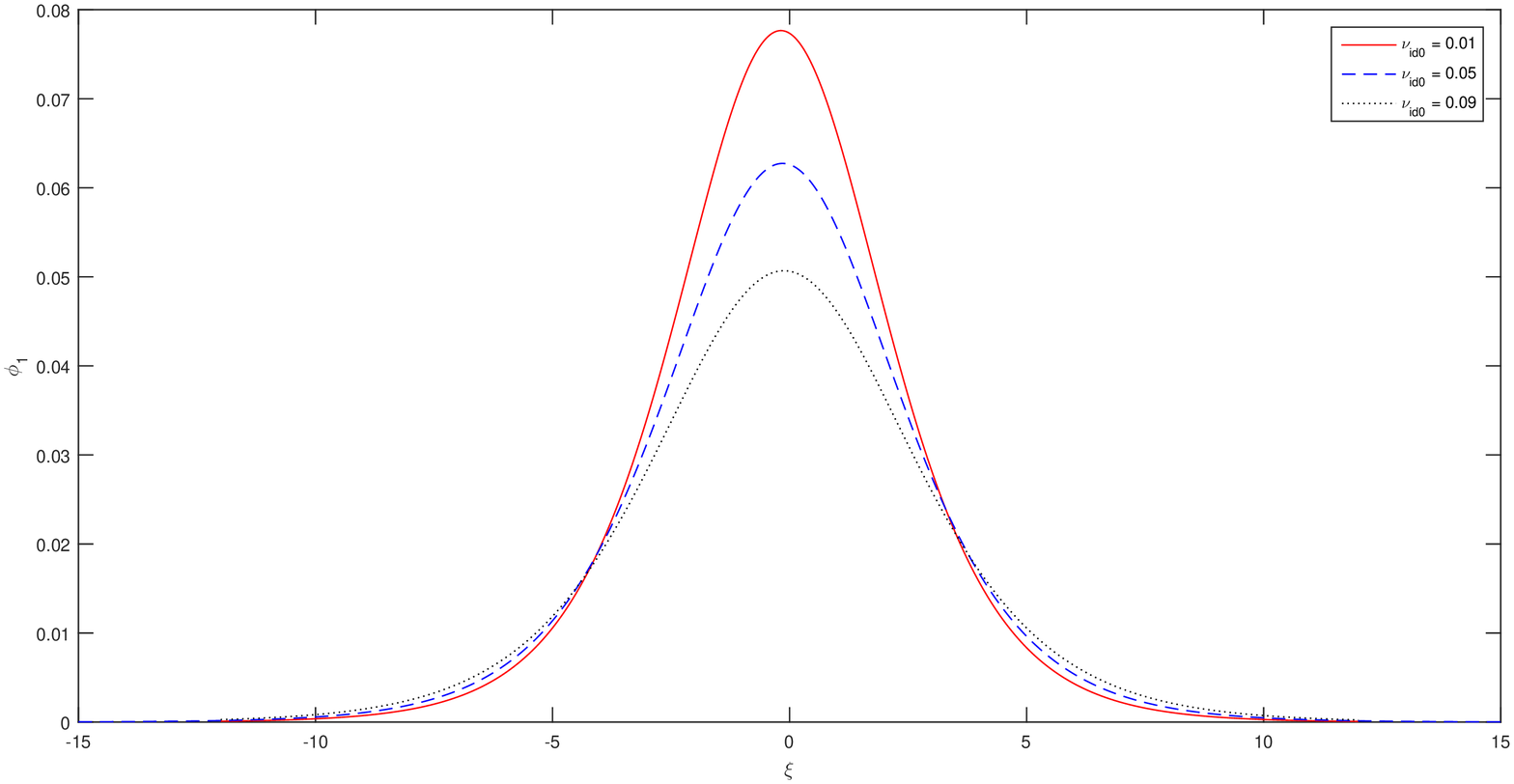}
\caption{Variation of the solitary wave of the non-planner damped KdV equation (\ref{p4eq12}) for different values of $\nu_{id0}$ with $M=0.1,\tau=-7,\mu=0.1,\kappa=1.8,\tau_0=-14$ and $\gamma=2$.}\label{f6}
\end{figure}
Our main interest to see the role of the damping effect as well as the cylindrical and spherical geometry upon the new analytical solitary wave solution, which may lead to close to our objectives. To emphasis the damping effect upon the solitary wave of the non-planner damped KdV equation, figure 5 and figure 6 are drawn for various values of damping parameter $\nu _{nd0}$ taking rest of the parameters as $M=0.1,\tau=-7,\mu=0.1,\kappa=1.8,\tau_0=-14$ for both  cylindrical and spherical geometry. System experiencing more collisions between dust grains and other plasma species leads to the increment of the damping effect. Increment of damping parameter bit by bit must be followed by the change of the solitary wave solutions. When collisions of dust grains and plasma species are more effective due to increase of temperature, pressure etc. Both figures i.e figure 5 and figure 6 delimitate that the solitary wave structures are flattened due to the increase of damping parameter. Delicate balance between nonlinearity and dispersion generates the solitary wave structures. Actually, nonlinearity steepens the solitary wave structures and dispersion spread it away. Here we observe that the damping may disperse the solitary wave structures.  Clearly it is seen that the damping parameter $\nu _{nd0}$ significantly modify the solitary waves solution which we have derived analytically.\\
\begin{figure}
\centering
\includegraphics[scale=0.5]{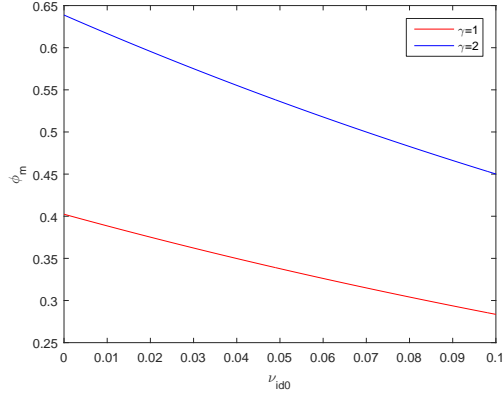}
\caption{Variation of the amplitude of the solitary wave of the non-planner damped KdV equation (\ref{p4eq12}) for a range of $\nu_{id0}$. Other parameters are same as those in figure \ref{f1}.}\label{f7}
\end{figure}

\begin{figure}
\centering
\includegraphics[scale=0.3]{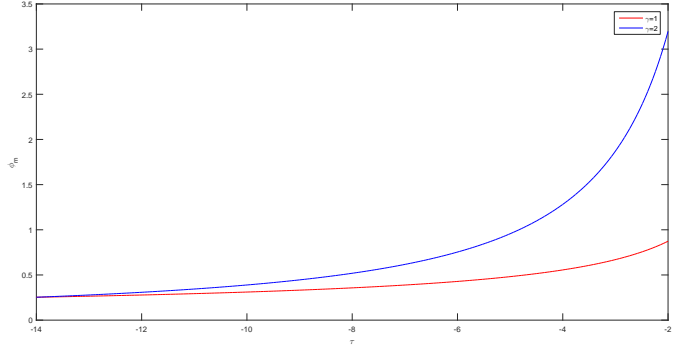}
\caption{Variation of the amplitude of the solitary wave of the non-planner damped KdV equation (\ref{p4eq12}) for a range of $\tau$. Other parameters are same as those in figure \ref{f1}.}\label{f8}
\end{figure}
Figure 7 is plotted to show the variation of the amplitude of the solitary wave of the non-planner damped KdV equation against the damping parameter within the range $0.01\leq \nu _{nd0}\leq 0.1$. The collisional effect strictly diminishes the amplitude of the solitary wave when it is gradually increased in both cylindrical and spherical geometry. On the other hand figure 8 depicts the change of width with relative to the damping parameter. As damping effect arises, the amplitude becomes lesser and width increases. This recognises the well known amplitude width relation. \\
\begin{figure}
\centering
\includegraphics[scale=0.5]{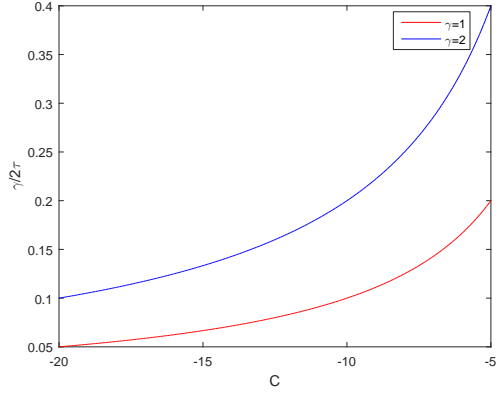}
\caption{variation of $\nu_{id0}$ Vs C.}\label{f9}
\end{figure}
In the above discussion we have shown that the parameter's significant effect on the derived analytical solution of the solitary wave solution step by step and these are generally studied in plasma dynamics. But our main focus of this article is the damping effect as well as the geometrical prospects altogether. It is observed that in our newly derived damped nonplanar KdV equations $\frac{\gamma }{2\tau }$ comes due to nonplanar geometry and $C $ is the contribution of damping effect. Analytically, in absence of collision of dust grains and other plasmas species, i.e. $C=0$ and the equation is nothing but a nonplanar KdV equation and the solution of nonplanar KdV equations is matched with our newly derived analytical solution. As the collision effect of plasma particles can not be ignored so we deal with the damping effect and collisional effect jointly. There is a very strong and direct relation between the damping parameter and the nonplanar geometry frame which is  observed from the derived equations. Using the relation $C+\frac{\gamma }{2\tau }\neq 0$ we can observe solitary waves structures in nonplanar geometrical frame in stead of planar geometrical frame.
\section{Conclusions}
It can be concluded that to study the effect of the dust-ion collision on solitary wave in dusty plasma  in unbounded geometry, the investigation have been made in the framework of planner evolution equation with with damping term. In this paper we extend those work in non planner geometry. Accordingly damped cylindrical KdV equation and damped sherical KdV equation are derived using RPT.
Study of dynamics of plasma within bounded nonplanar geometry in more nearer to reality. Experimental equipments are bounded cylindrical or spherical, it is often difficult to observe through unbounded planar geometrical frame. In point of fact, the solitary wave structures are found when we want to observe that through some moving frame of reference which is followed by the stretching of independent variables in the RPT. Figure 9 is plotted to illustrate the dependency of the damping and the nonplanar frame of reference. Along the red(blue) line the cylindrical(spherical) geometrical effect and damping effect altogether reduce damped nonplanar KdV equations into simple KdV without withdrawing the damping effect. Similarly, one can find their required range in nonplanar geometry where solitary wave structures may be found. This is one of the way of getting the suitable range of cylindrical and spherical solitary waves using the planar solitary wave solution with damping in permissible range, which is not yet been reported in plasma literature. This way will strengthen the relationship between theory and experiments.

\end{document}